 \documentclass [12pt,a4paper      ]{article}
\usepackage{graphics}
\usepackage{times}

\DeclareFontFamily{OT1}{times}{}
\DeclareFontShape {OT1}{times}{m }{n }{ <-> ptmr }{}
\DeclareFontShape {OT1}{times}{bx}{n }{ <-> ptmb }{}
\DeclareFontShape {OT1}{times}{m }{it}{ <-> ptmri}{}
\DeclareFontShape {OT1}{times}{bx}{it}{ <-> ptmbi}{}
\usepackage{amsmath}
\usepackage{amsfonts}
\usepackage{amssymb}
\usepackage{latexsym}
\newcommand{\cl}{C \kern -0.1em \ell} % Clifford algebra
\newcommand{\CON}{\overline}          % quaternion (vector)conjugate

\newcommand{\VEC}{\vec{\kern +.1em[}} % VEC-tor left  bracket   |>.. 
\newcommand{\TOR}{\vec{\kern +.2em]}} % vec-TOR right bracket   ..>| 
\newcommand{\BRA}{\langle\kern -.2em\langle} % Dirac BRA      <<...| 
\newcommand{\KET}{\rangle\kern -.2em\rangle} % Dirac KET      |...>> 

\begin{document}

\title{\bf\vspace{-2.5cm} 
   Antimatter induced fusion and thermonuclear explosions\footnote{Full-length version of a paper contributed to the 4th Int. Conf. on Emerging Nucl. Energy Syst., Madrid, June 30/July 4, 1986. Published {\bf in} Atomkernenergie $\cdot$ Kerntechnik  (Independent Journal on Energy Systems and Radiation) {\bf 49} (1987) 198--203.}}
%                         ========================================

\author{{\bf Andre Gsponer and Jean-Pierre Hurni}\\
\emph{Independent Scientific Research Institute}\\ 
\emph{Box 30, CH-1211 Geneva-12, Switzerland}\\
e-mail: isri@vtx.ch\\}

\date{ISRI-86-06.4 ~~ \today}

\maketitle

\begin{abstract}

The  feasibility  of  using antihydrogen  for  igniting  inertial 
confinement   fusion   pellets   or   triggering   large-scale 
thermonuclear   explosions  is  investigated.    The  number   of 
antiproton  annihilations required to start a thermonuclear  burn 
wave in either $DT$ or $Li_2DT$ is found to be about $10^{21}/\kappa^2$,  where $\kappa$ 
is the compression factor of the fuel to be ignited. 
 
     In   the  second  part,   the  technologies  for   producing 
antiprotons  with  high energy accelerator systems and the  means 
for  manipulating and storing microgram amounts  of  antihydrogen 
are  examined.   While there seems to be no theoretical obstacles 
to  the  production  of  $10^{18}$ antiprotons  per  day  (the  amount 
required for triggering one thermonuclear bomb), the construction 
of  such a plant involves several techniques which are between  3 
and 4 orders of magnitude away from present day technology. 
 
     Considering  the financial and energy investments needed  to 
produce antimatter, applications will probably remain confined to 
the military domain.  Since antihydrogen-triggered  thermonuclear 
explosives  are very compact and have extremely reduced fallout, 
we  conclude that such devices will enhance the proliferation  of 
nuclear weapons and further diffuse the distinction between  
low-yield nuclear weapons and conventional explosives.

\end{abstract}

\section{Introduction}
%---------------------

Matter-antimatter  interaction  produces  more 
energy  per unit mass than any other means  of 
energy  production. For  example, proton-antiproton
  annihilation  releases  275  times 
more  energy in the form of kinetic energy  of 
charged  particles than nuclear fission or  $DT$ 
fusion.   This  energy  is released by  simple 
contact  of antimatter with matter so that, in 
principle,  no ignition energy is required  to 
start  the  reaction.   It  is  therefore  not 
surprising   that   the   concept   of   using 
antimatter  as  an energy source has  been  in 
scientific  literature for decades \cite{1,2}.

     Other  practical  applications of   antimatter
 are under consideration.  For  example, 
antimatter  propulsion  systems  \cite{3},    space 
based  power generators \cite{4},  directed  energy 
weapons \cite{4},\footnote{Directed  energy weapons applications may 
include the projection of plasma 
jets, X-ray or gamma-ray laser pumping, and antimatter beams.} and cancer therapy \cite{5,6}. Finally, both 
\emph{Edward  Teller} \cite{7,8,9,10} 
and \emph{Andrei Sakharov} \cite{11}, 
the   key   scientists   in  charge   of   the 
development of the H-bomb in their  respective 
countries,  show in their published scientific 
works  a  big  interest  in  the  annihilation 
properties of antimatter,  the nuclear process 
that  after fission and fusion could lead to a 
third generation of nuclear bombs. 

     This   paper   is   a   summary   of    a 
comprehensive assessment of the feasibility of 
producing  large quantities of antiprotons and 
using  them for igniting  inertial-confinement 
fusion   pellets  or  triggering  large-scale 
thermonuclear explosions \cite{12}.  In 
sections 2 to 6  we evaluate the number of 
antiprotons  needed to start a thermonuclear detonation  wave  in 
either $DT$ or a $Li_2DT$ mixture.  In Sections 7 to 11 we examine the 
problems  of producing,  collecting,  cooling,  manipulating  and 
storing the required amounts of antiprotons and antihydrogen.

\section{Matter-antimatter annihilation}
%---------------------------------------
 
When  a particle meets it's antiparticle  they 
annihilate  and the energy equivalent to their 
total  mass ($2mc^2$) is converted  into  various 
new particles and kinetic energy \cite{13}.   In the 
case  of proton-antiproton annihilation,  many 
different reaction channels are possible, each 
resulting  in the production of  a   different 
number  of charged and neutral  particles.   A 
 good  approximation is that three charged  and 
two neutral pions are produced on the average.  
Since   neutral   pions  quickly  decay   into 
photons,  the typical $p\CON{p}$ annihilation  process 
is as follows: 
\begin{equation}\label{EQ-1}
   p + \CON{p} \rightarrow 3 ~ \pi^\pm + 2 ~ \pi^0
               \rightarrow 3 ~ \pi^\pm + 4 ~ \gamma,
\end{equation}
where $E_\pi^\pm = 236$~MeV and $E_\gamma = 187$~MeV.
An  antiproton  can  also  annihilate  with  a 
neutron,   in  which  case  mostly  pions  are 
produced again,  in numbers,  on the  average, 
similar to $p\CON{p}$ annihilation. 

     Antiprotons,  antineutrons  and positrons 
can  combine to  form  antinuclei,  antiatoms, 
antimolecules.   Annihilation  occurs when the 
two kinds of matter come sufficiently close to 
one other.   Even at some distance,  a neutral 
atom and  a neutral antiatom will attract each 
other  by  van  der Waals forces  \cite{10,13}.   As  a 
consequence,   storage   of \emph{antiatoms} in   a 
container  made  of matter  is  impossible  in 
general.   However, there may exist metastable 
states of \emph{antiprotons} in normal matter \cite{14}, 
and $\CON{p}$'s may possibly 
be stored in superfluid helium \cite{14},  a speculation encouraged by 
the  fact  that helium is the  only  atom  which,  theoretically, 
cannot capture a low energy antiproton \cite{15}.

\section{Plasma heating with antiprotons}
%----------------------------------------
 
When  a  $\CON{p}$ annihilates in a  hydrogen  plasma, 
essentially  all  the annihilation  energy  is 
radiated  in the form of very energetic  pions 
and photons.  At solid hydrogen densities, the 
mean free path of the 187-MeV photons is 25~m, 
so  that  they  will not loose energy  in  the 
plasma.   However,  the three 236-MeV  charged 
pions  will  loose energy by multiple  Coulomb 
interactions  with  the electrons  at  a  rate 
approximately given by: $dE/dx = 0.52$~MeV/cm in 
solid $H_2$ or $DT$ and 2.06~MeV/cm in $Li_2DT$.
 
     If  we now assume that annihilation takes 
place  at the center of a sphere,  the  energy $dW$ 
deposited  within a radius $R=1$~cm is  only   
1.5~MeV  out  of the total 1876~MeV  annihilation 
energy.   There  are however several ways   to 
improve  energy  deposition,  and thus  plasma 
heating.   Firstly,  the fuel to be heated may 
be compressed by a factor $\kappa$,  $dE/dx$ will  then 
be multiplied by $\kappa$,  and thus $dW$ by $\kappa^{2/3}$.  But 
compression requires energy.  Secondly,  fuels 
such  as $Li_2DT$,  which contain more electrons, 
have a proportionally larger $dE/dx$.   However, 
their  thermonuclear ignition  temperature  is 
also higher.   Finally,  annihilation may take 
place with a nucleus. 
 
     When  a $\CON{p}$ annihilates with a nucleon from 
a nucleus,  because of the Fermi motion of the 
annihilated nucleon,  the nucleus will  recoil 
with an energy of about 20~MeV.   Furthermore, 
each   of  the  5  annihilation  pions  has  a 
probability of colliding with the rest of  the 
nucleus.   Hence,  the  average  total  energy 
deposition in a sphere is 
\begin{equation}\label{EQ-2}
 dW = \nu \frac{dE}{dx} R + \epsilon ,
\end{equation}
where  $\nu = 3$ is the number of charged pions and    
$\epsilon$ the local energy deposition by the recoiling 
nucleus    and   the   various    pion-nucleus 
interaction debris. 

     In  the case of $\CON{p}$ annihilation with  
deuterium  or tritium  $\epsilon$ is approximately 12~MeV 
on the average,  about half of the energy
corresponding to the Fermi momentum.   
With  heavy nuclei there have been  many 
theoretical  speculations  in the  absence  of 
measurements.   The  first of these was introduced  
by \emph{Duerr} and \emph{Teller} \cite{9}, who  speculated 
that  an antiproton would find a  very  strong 
(900~MeV)  attractive potential when  getting 
close to a nucleus.   More recently \cite{16},  Los 
Alamos  scientists have calculated that annihilation  
in carbon would result in the  local 
energy  deposition of about 100~MeV.   Recent 
measurements  at CERN show that it is in  fact 
only  33~MeV in carbon \cite{6}, and approximately 55~MeV 
in silicon \cite{17}.   Low energy  $\CON{p}$'s  
annihilate  mostly at the surface  of  nuclei, 
and  thus  local energy deposition  follows  a 
$A^{2/3}$ dependence on atomic weight.   In effect, 
the   CERN   data  is  compatible   with   the 
expression : 
\begin{equation}\label{EQ-3}
    \epsilon \approx 6.4 A^{2/3}  ~~~ \text{[MeV]}.
\end{equation}
Hence,  for $\CON{p}$ annihilation in $H_2$, $DT$ or $Li_2DT$,       
$\nu$  is  always  about 3 and $\epsilon$ is  approximately 
equal to 0, 12 or 22~MeV respectively. 
 
\section{Thermonuclear burn of a particle-antiparticle plasma}
%------------------------------------------------------------- 
 
A matter-antimatter plasma is obtained if some 
initially stable particle-antiparticle mixture 
is suddenly ignited.  The annihilation rate of 
two interacting species, with number densities 
$n$ and $\CON{n}$, is 
\begin{equation}\label{EQ-4}
        \frac{d\CON{n}}{dt} = - n \CON{n} \langle \sigma v \rangle ,
\end{equation}
where  $\langle \sigma v \rangle$ is the annihilation reaction rate 
averaged over the Maxwell distribution. Fig.~1 gives $\langle \sigma v \rangle$ for $e\CON{e}$ and $p\CON{p}$ plasmas.  
 
\begin{figure}
\begin{center}
\resizebox{12cm}{!}{ \includegraphics{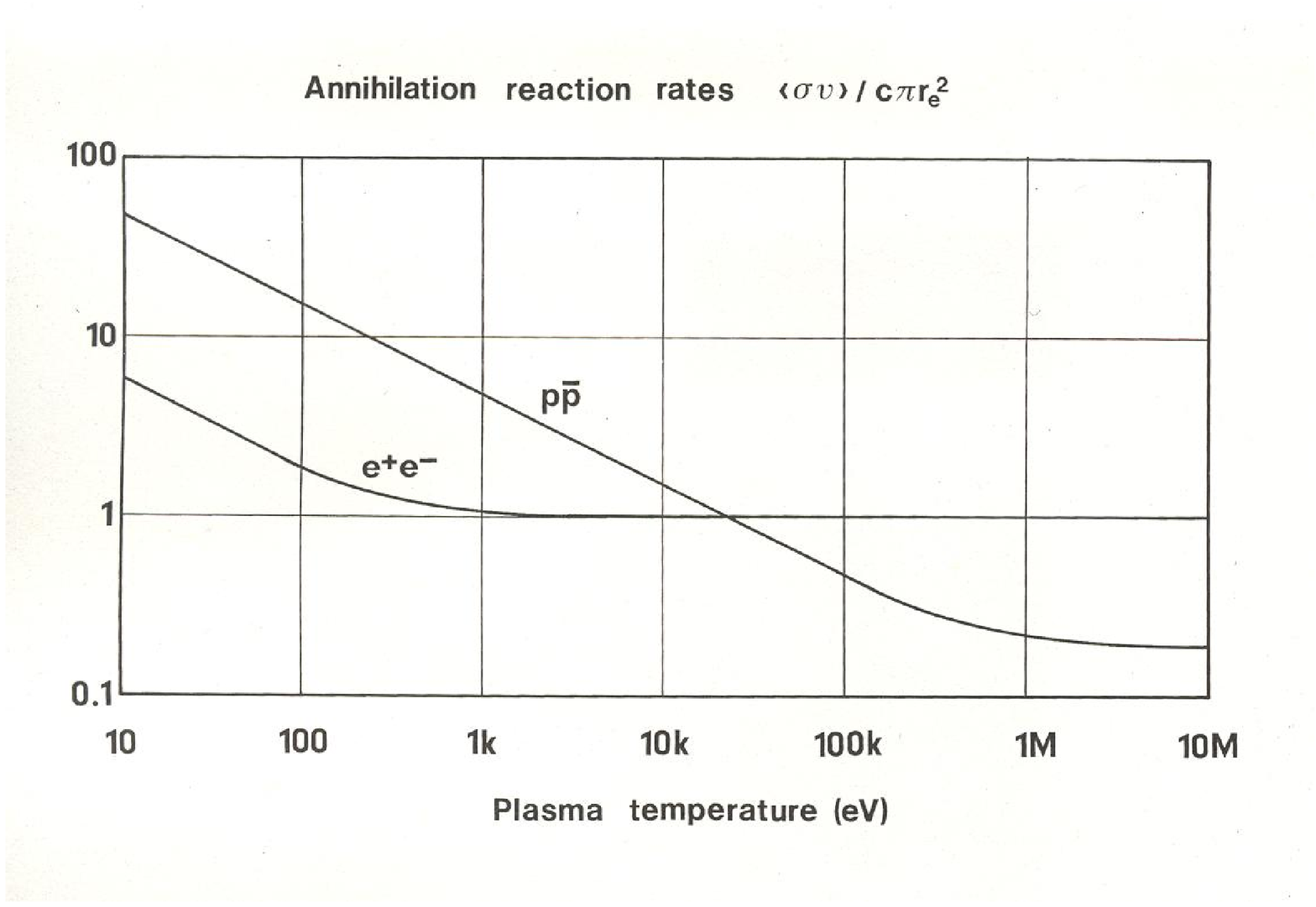}}
%...................................................
\end{center}
\caption[Annihilation reaction rates]{Electron-positron and proton-antiproton  annihilation reaction rates averaged over the Maxwell velocity distribution.}
\end{figure}

In a $H - \CON{H}$ plasma,  equation \eqref{EQ-4} holds for  both 
protons  and  electrons  with  $n = \CON{n}  =  n_0  = \rho \, \mathcal{N}_A/2$
initially. Hence, for a given temperature 
\begin{equation}\label{EQ-5}
     n = \frac{n_0}{1+t/\tau}  ~~~  ~~~ \text{with} ~~~ ~~~
  \tau =\frac{2}{n_0\langle \sigma v \rangle} .
\end{equation}
If  we assume $T = 20$~keV,  $\langle \sigma v \rangle$ is  approximately
  the  same for both $e^+ e^-$ and $p\CON{p}$  
annihilation.   Thus  the electron and the  proton 
populations deplete  at the same rate,  with a  
time constant  of  5~ns  for   $\rho = 0.07$~g/cm$^3$.

\section{Annihilation in a matter-antimatter boundary layer}
%----------------------------------------------------------- 
 
When  matter and antimatter come into contact, 
annihilation   primarily  takes  place  in   a 
boundary   layer   in  which   particles   and 
antiparticles are mixing.   The thickness   of 
this  matter-antimatter plasma is of the order 
of  the  antimatter mean-free-path in  matter, 
i.e.,  $(3n\sigma)^{-1}$.   A first approximation,  
assuming  that  whenever an  antiparticle  
penetrates  into  the boundary layer it  instantly 
annihilates,   is  an  annihilation  rate  per 
element  area  given by the  total  number  of 
antiparticles impinging on that surface.  From 
the Maxwell velocity distribution one gets 
\begin{equation}\label{EQ-6}
       \frac{d\CON{N}}{dS\,dt} = -\CON{n}c \sqrt{\frac{kT}{2\pi mc^2}}  .
\end{equation}
The $e^+$ annihilation rate is thus  $\sqrt{m_p/m_e}  \approx  43$ 
times the $\CON{p}$ annihilation rate.  However, since 
the $\CON{H}$ plasma Debye length is much smaller than 
the  boundary layer thickness,  plasma  charge 
neutrality  insures  that the antimatter  flow 
rate is determined by the slowest annihilation 
rate.  Therefore,  if  $\CON{H}$'s interact  with  the 
walls of a closed cavity, annihilation results 
in  an  overall  decrease  of  the  antimatter 
density within the cavity.
 
     Let  us now take the case of a sphere  of 
solid  antihydrogen  that is suddenly  put  in 
contact  with a collapsing spherical shell  of 
compressed $DT$ (see Fig.~2).  To solve Eq.~\eqref{EQ-6} one has to 
calculate  the  increase in the $\CON{H}$  plasma  internal 
energy by the pions and other particles from $\CON{p}$ 
annihilation in the surrounding $DT$: 
\begin{equation}\label{EQ-7}
    dW = -dN \frac{1}{2}
          \Bigl( \nu \frac{dE}{dx} + \frac{\epsilon}{\lambda} \Bigr)
           \frac{4R}{\pi} \frac{N}{N_0} ,
\end{equation}
where $\lambda  = 3$~cm is the approximate range of the  
20-MeV recoil protons from $\CON{p}$ annihilation  in 
$DT$,  and $N$ (initially equal to $N_0$) the  number 
of $\CON{H}$ atoms.  For hydrogen  $dW = 3 N k dT$, we get 
a system of equations for the $\CON{H}$ plasma density 
and  temperature.   If  annihilation  is  much 
faster  than  the collapse of  the  cavity, $R$ remains
constant and the solution of Eqs.~\eqref{EQ-6} and \eqref{EQ-7} is 
\begin{equation}\label{EQ-8}
    T= T_1 \tanh^2 (t/\tau_a) ,
\end{equation}
and
\begin{equation}\label{EQ-9}
    N = N_0 \Bigl(1 -  \tanh^2 (t/\tau_a) \Bigr) .
\end{equation}
For $N_0 = 10^{18}$,  which corresponds to $R =  0.02$~cm,  
we  find  $T_1 = 19$~keV and $\tau_a  = 0.25$~ns. 
Thus, in about  $2\tau_a  = 0.5$~ns, over 90\% of the 
antihydrogen  in  the sphere  is  annihilated.  
This  time  constant is  compatible  with  the 
requirements  of instantaneous  thermalization 
and inertial confinement of the plasma.

\begin{figure}
\begin{center}
\resizebox{8cm}{!}{ \includegraphics{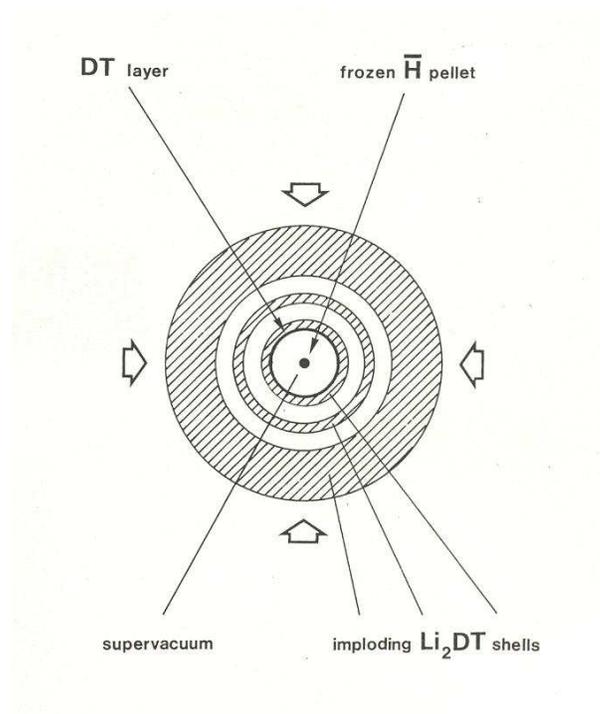}}
%...................................................
\end{center}
\caption[Ignition  of a spherical thermonuclear detonation wave]{Ignition  of a spherical thermonuclear detonation wave in   $Li_2DT$.  A series of concentric shells are imploded by chemical explosives 
or  by other means.   When the innermost shell gets into  contact 
with  the levitated antihydrogen  pellet,  annihilation  produces 
sufficient  energy  to trigger a thermonuclear burn wave  in  the 
bulk  of  the   $Li_2DT$  fuel.    The  multishell  structure  avoids 
excessive preheating of the antihydrogen pellet during implosion.}
\end{figure}

\section{Antiproton triggered thermonuclear detonation wave}
%-----------------------------------------------------------
 
The   most   efficient   way  to   trigger   a 
thermonuclear explosion is probably to start a 
thermonuclear  detonation  wave  in  $Li_2DT$  by 
collapsing a hollow sphere of that material on 
a tiny spherical pellet of solid antihydrogen (Fig.2). 

     In   the  spark  model  of  thermonuclear 
ignition \cite{18},  an  outgoing  spherical  detonation 
wave  starts  if :  (a) a critical  amount  of 
energy  $E_c$ is deposited in the center  of  the 
sphere  (the  "spark" region) and (b)  if  the 
temperature  within this volume is higher than 
a    critical   temperature   $T_c$.     Without 
compression,   one  has  $E_c = 5 \times 10^{25}$~keV   and  
$T_c  = 4$~keV  for  solid $DT$ ,   and $E_c = 3 \times 10^{26}$~keV 
and $T_c = 13.6$~keV for  $Li_2DT$.   However, 
for   a  compressed  thermonuclear   fuel   at 
temperature $T_c$,  the critical energy decreases 
with the square of the compression factor $\kappa$. 

     The number $N$ of $\CON{p}$ annihilations necessary 
to  induce  a thermonuclear burn wave  can  be 
estimated by supposing that annihilation takes 
place  at  the  center  of the  sphere  to  be 
ignited.   Thus,  from equation \eqref{EQ-2}, condition 
(a) is satisfied if 
\begin{equation}\label{EQ-10}
  E_c/\kappa^2 = N \Bigl( \nu \frac{dE}{dx} \kappa R_s + \epsilon \Bigr) .
\end{equation}
Since the pions originate from the center, the 
temperature   in   the  fuel  goes  as   $1/r^2$.  
Therefore,  for  simplicity,  we require  that 
condition  (b)  is satisfied for  the  average 
temperature within the critical volume.  Thus 
\begin{equation}\label{EQ-11}
E_c/\kappa^2 = \frac{3}{2} \frac{z}{a} \kappa \rho N \frac{4\pi}{3} R_s^3 k T_c,
\end{equation}
where $z$ and $a$ are respectively equal to 2  and 
2.5 for $DT$,  and 6 and 9.5 for  $Li_2DT$.   Taking     
$\kappa =  30$,  a  modest  compression  factor,  and  
solving Eqs.~\eqref{EQ-10} and \eqref{EQ-11} 
for $N$ and the spark radius $R_s$, 
one finds $N = 3 \times 10^{18}$ and $R_s = 0.09$~cm for $DT$, 
and $N = 6 \times 10^{18}$ and $R_s = 0.07$~cm for  $Li_2DT$.  
However,  because  of some of the  simplifying 
assumptions   made,   these  results  may   be 
somewhat pessimistic.   Hence,  we will assume 
that $10^{18}$ $\CON{p}$'s are sufficient to trigger  the 
thermonuclear  explosion of compressed  $DT$ or 
$Li_2DT$ pellets. 

     For  thermonuclear explosions in  the  kiloton 
range,  chemical  explosives  may be  used  to 
implode  the  $Li_2DT$  shells.   For  low  yield 
explosions such as in X-ray laser pumping  or
inertial-confinement fusion (ICF), 
compression factors higher than 30 can be 
achieved using magnetic compression,  beams or 
other techniques.  However, antiproton induced 
fusion  will remain an attractive  alternative 
to  normal ICF only if the compression  factor 
is kept relatively small,  i.e., less than 300, 
giving a number of $\CON{p}$'s of the order of $10^{16}$.

\begin{figure}
\begin{center}
\resizebox{8cm}{!}{ \includegraphics{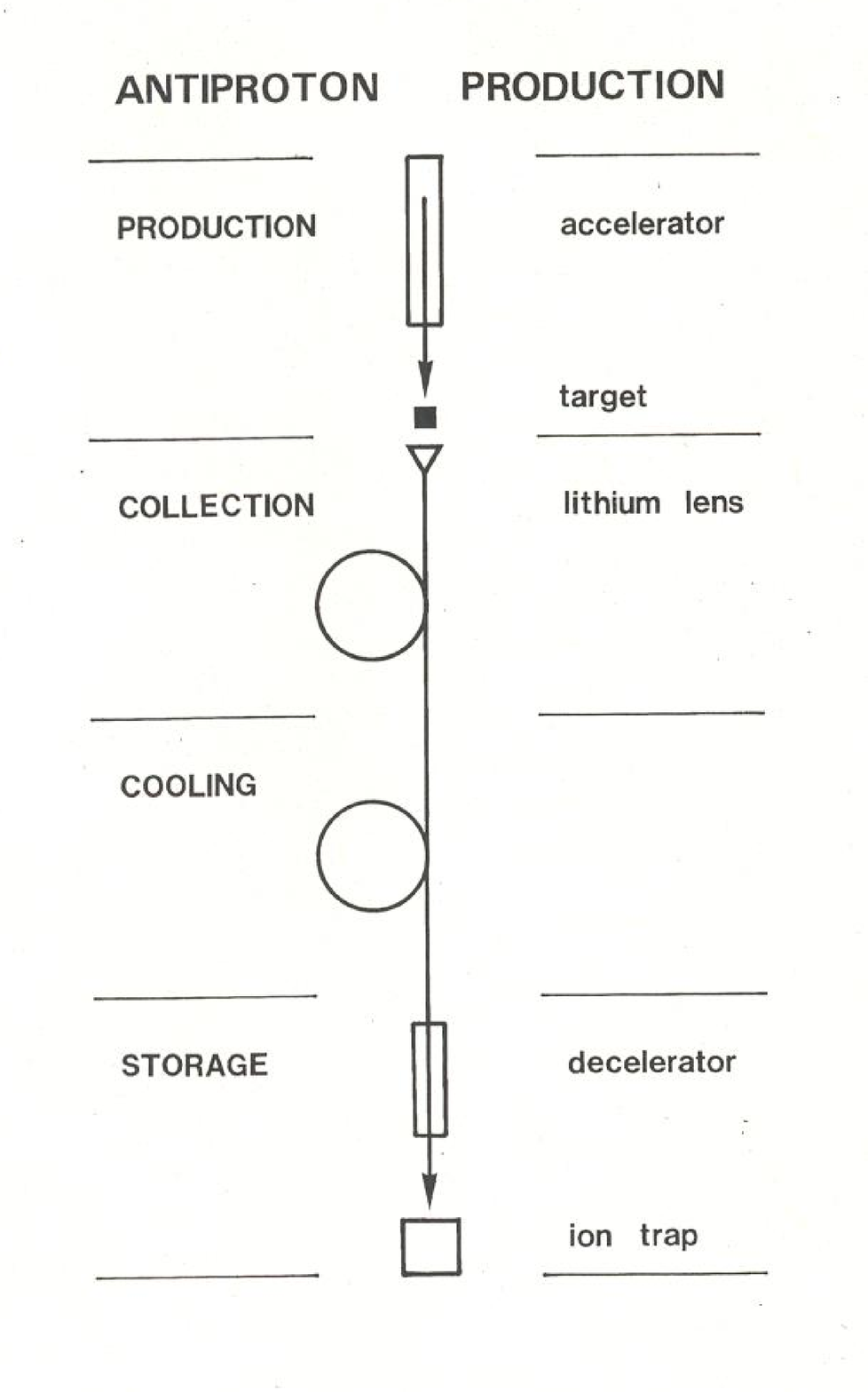}}
%...................................................
\end{center}
\caption[Main antiproton production steps]{The four main steps of present-day technology antiproton production.}
\end{figure}

\section{Antiproton production}
%------------------------------
 
There are 4 main steps from $\CON{p}$ production in high energy  particle 
collisions, to the manufacture and storage of solid $\CON{H}$ (Fig.~3).  
In current systems, antiprotons are produced when protons of high 
enough  energy (over 6~GeV) are fired into a target.   These  $\CON{p}$'s 
emerge  with  a  wide variety of energies and a  whole  range  of 
angles.   This  very broad beam of $\CON{p}$'s can be characterized by  a 
very high temperature,  of the order of 100's of MeV.  The second 
step is to collect as wide a range of antiprotons as possible and 
to  start concentrating them in velocity and angle while  storing 
them  in a first high energy storage ring.   The third step is to 
accumulate  them  in a second ring while  continuously  "cooling" 
them until they all have the same velocity and  angle.   Finally, 
when  the  $\CON{p}$'s are cold enough,  they can be decelerated to  zero 
velocity   and   combined   with  positrons   to   form   neutral 
antihydrogen.   In  this Section and  the following  three ones we 
examine   the  state-of-the-art  in  these  techniques  and   the 
possibility of using them for large scale antimatter  production, 
i.e., $10^{13}$ $\CON{p}$/s ($10^{-6}$~g of $\CON{H}$ or $10^{18}$ $\CON{H}$ atoms per day). 
 
     The  only antiproton factory in operation today is  at  CERN 
near Geneva. It produces $2 \times 10^6$ $\CON{p}$/s at the output of it's
 storage-cooling ring.  By 1987, this system will be upgraded to produce $2 \times 10^7$ $\CON{p}$/s \cite{19}.  At Fermilab, near Chicago, an antiproton 
source of the  same intensity is under construction \cite{20}.   Antiprotons are 
also produced in the USSR \cite{1} where there are plans for a  system 
that  will  permit storage of $10^8$ $\CON{p}$/s \cite{21}.   However,  the  most 
ambitious  project  is at Los Alamos where the $\CON{p}$  flux  from  the 
target is expected to be 100--200 times that of CERN \cite{22}. 
 
     The  economic  feasibility of an antiproton factory  depends 
crucially on the accelerator system's transformation  coefficient 
of  electricity into antiproton rest mass.   Since the number  of 
antiprotons produced increases logarithmically with the collision 
energy,  there is a broad  optimum at 120~GeV, precisely the beam 
energy of the Fermilab $\CON{p}$ source.   However, compared with a fixed 
target system,  the use of a particle-particle collider \cite{23} is a 
much  more  efficient means for high yield  particle  production.  
With this method the optimum corresponds to a (16+16)~GeV collider.  
Such   a  collider  could  be  built  at  Los  Alamos  where  the 
construction  of  a  high  intensity  8 to 45~GeV  synchrotron  and 
possibly a future colliding beam facility \cite{22,24} are projected.  
 
     Colliding  beams  of heavy ions \cite{25} may  be  an  attractive 
alternative.  Indeed,  a  heavy  ion  collider  of  the  required 
luminosity  might be easier to build than a proton collider,  and 
in very high-energy-heavy ion-collisions, one expects an enhanced 
production of antiparticles such as antiprotons \cite{26}.

\section{Antiproton collection}
%------------------------------

In  fixed  target systems,  both at CERN and Fermilab,  a  Soviet 
designed  lithium  magnetic lens \cite{27} is used to capture  a  wide 
spread  of  $\CON{p}$'s as they are produced at the target and  to  focus 
them on the aperture of the $\CON{p}$ collection channel (Fig.~3).   A 
plasma lens \cite{19,28} could be used instead to improve the  angular 
acceptance.  However,  it  is  more  important  to  increase  the 
momentum  acceptance  which is only 1 to 2\% in present day  systems.  
For that purpose, a linear debuncher between the production point 
and  the first $\CON{p}$ storage ring could be used \cite{29}.  Together  with 
other  possible improvements,  the overall collection  efficiency 
could be as high as 0.05. To produce $10^{13}$ $\CON{p}$/s,  assuming an 
electric power efficiency of 25\%, the current of a 120~GeV beam would 
be 1~mA and the power load for the accelerator about 450~MW. 
 
     To collect the $\CON{p}$'s from a colliding beam source, there is an 
advantage  in having a small asymmetry in the two beam  energies, 
for example 14 and 18~GeV.  The center of mass energy would still 
be  very  close to optimum but the $\CON{p}$'s produced at the  threshold 
would go precisely in the direction of the fast beam, and with an 
energy  equal  to  the  difference of  the  beam  energies  \cite{30}. 
Assuming  again  a collection efficiency of  0.05,  the  required 
luminosity for a proton-proton collider would be $6 \times 10^{40}$ for  each 
of the~8 interaction points as in Fig.~4.  The total beam current 
for  supplying 14 and 18~GeV protons to the collider is then  0.6 
mA and the power load  50~MW. These numbers are quite close to 
the state-of-the-art. For example, the present 0.8~GeV Los Alamos 
accelerator  normally  runs with a current of up to 0.9~mA,  and 
there  is a proposal to accelerate 0.17 mA of that beam to 8  GeV 
and  as  much as 0.07~mA up to 45~GeV  \cite{22}.  But   building  the 
desired collider will be a much more difficult task \cite{25},  unless 
a big $\CON{p}$ production enhancement in heavy ion collisions is found.

\section{Antiproton cooling}
%--------------------------- 
 
Cooling aims at reducing the angular  and energy spread of a beam 
circulating  in a storage ring.  There are two basic  techniques: 
electron  cooling which was pioneered in the Soviet Union \cite{21,31} 
and stochastic cooling which has been invented at CERN \cite{32}.   In 
many  respects electron and stochastic cooling are  complementary 
\cite{33}. The efficiency of electron cooling is best for the cold and 
stochastic  cooling for the hot beams.   This suggests  combining 
pre-cooling with stochastic and final cooling with electrons. 
 
     Stochastic  cooling systems based on present techniques  are 
capable of cooling as many as $10^8$ to $10^{10}$ $\CON{p}$/s \cite{32}. 
Even with 8 systems working in parallel this is short by 2 to 4 
orders of magnitude of  being able 
to produce $10^{13}$ $\CON{p}$/s.  The only solution  known  at 
this  time  to  go beyond this limit is to use  multiple  cooling 
rings.  If each of these rings is fed by a different  collection 
channel,  the theoretical improvement in the overall cooling rate 
is  $n \ln n$, about 23, for 10~rings working in parallel. 
 
     The  main advantage of electron cooling is that it does  not 
suffer any intrinsic particle number limitation   \cite{21,33}.   But, 
unlike  stochastic  cooling,  electron cooling times  are  strong 
functions  of  the  $\CON{p}$  beam  momentum  and  $\CON{p}$ beam  temperature.  Furthermore,  electron  cooling  has never been tested with  more 
than $10^9$ particles \cite{33}.   Nevertheless, if sufficient debunching 
and  precooling  can  be achieved,  electron  cooling  should  be 
capable  of handling a rate of $10^{12}$ $\CON{p}$/s or higher.  This  is  why 
most  cooling  research in both the USA and the USSR has  
concentrated on electron cooling. A conceivable system would consist of 
three rings as in Fig.~4.   Then, if a combination of debunching, 
stochastic pre-cooling and electron cooling could cool $10^{12}$  $\CON{p}$/s,  
our problem would be solved.   This is by no means a simple task, 
but there does not appear to be any fundamental obstacles.

\begin{figure}
\begin{center}
\resizebox{12cm}{!}{ \includegraphics{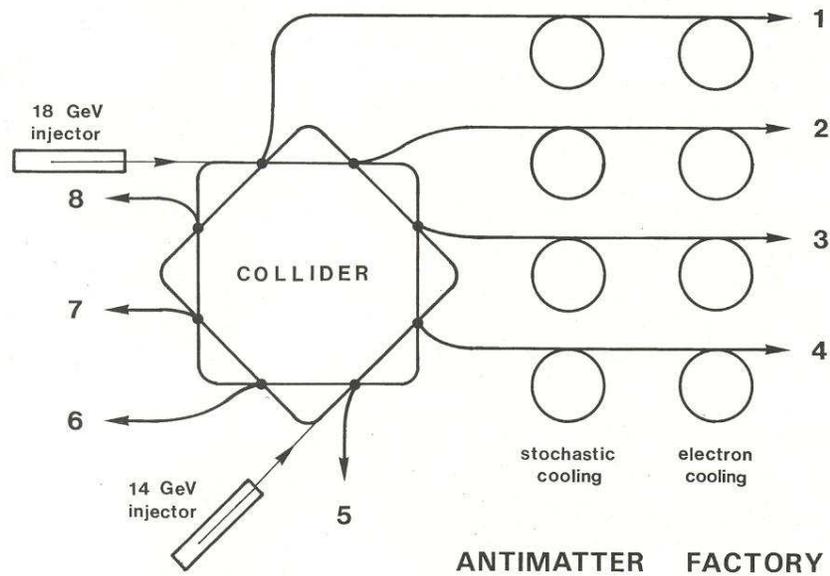}}
%...................................................
\end{center}
\caption[Collider  based  antiproton  factory]{A  possible  design  for a  collider-based  antiproton  factory.  
Antiprotons  are  produced at the eight interaction points  where 
the  high  energy ($\approx$16~GeV/nucleon) proton or heavy  ion  beams 
collide.   The  antiprotons are collected,  cooled and  processed 
into  antihydrogen  by eight systems working  in  parallel.}
\end{figure}

\section{Antiproton storage}
%--------------------------- 

The  only  antimatter storage technique proven today is  that  of 
storage rings \cite{34,35}.   For practical applications, it is 
necessary  to find more permanent means for storage,  and for ease  of 
handling, if possible in solid form.  This problem has been 
studied extensively in the conceptual design of antimatter spacecraft 
propulsion systems \cite{3}.   Many different techniques are  feasible 
in principle, but they still have to be tested experimentally. 
 
     In any event,  the first thing is to decelerate the $\CON{p}$'s down 
to  a  few  eV.   For that purpose,  the  most  promising  method 
involves  the  use  of  a radio-frequency quadrupole  \cite{36}  as  a 
decelerator  from 5~MeV (or more) down to approximately 100~keV, 
and  to catch the $\CON{p}$'s in a ion trap in which they are  cooled  by 
resistive  damping of image currents or by electron cooling \cite{14}.  
In  an ion trap (also called Penning trap) the density of  stored 
$\CON{p}$'s  can  reach  $10^{11}$ $\CON{p}$~cm$^{-3}$ \cite{37}.  The 
lifetime of  the  $\CON{p}$'s  is 
primarily  limited by annihilation after capture by the  residual 
gas  atoms.   Pressures lower than $10^{-15}$~Torr and  liquid  helium 
temperatures are required to keep the loss rate below $10^{-6}$ s$^{-1}$.

\section{Antihydrogen production and storage}
%--------------------------------------------

Much higher storage densities are possible if the antiprotons are 
combined   with  positrons  to  form  neutral  $\CON{H}$.  Antihydrogen 
formation is quite difficult \cite{38}.   However, since positrons are much 
easier to produce and cool than $\CON{p}$'s, large scale $\CON{H}$ production
is certainly feasible at a cost that would be marginal compared with 
the investment necessary for a full-scale $\CON{p}$ production plant. 
 
     Once neutral hydrogen has been formed,  it has to be further 
slowed and cooled.  Storage rings may be used to store $\CON{H}$, but for 
this   purpose,   cooling  and  subsequent  manipulation,   laser 
techniques are probably better.   One method is called  resonant 
radiation  cooling  and  capture \cite{39} which can also be  used  to 
create a trap for the atomic antihydrogen \cite{40}. 
 
     If  atomic $\CON{H}$ is transformed into molecular $\CON{H}_2$, it can be 
cooled  to  very  low temperatures where it will  assume  the  low 
energy parahydrogen state.   Since this molecule is  diamagnetic, 
it  can  be  directed to the storage container  by  hexapole-type 
magnetic field channels that have a zero field at the center.  At 
temperatures  below 14~K,  the $\CON{H}_2$ molecules can then condense  to 
form solid antihydrogen pellets which can be stored using  either 
magnetic, electrostatic or laser levitation techniques \cite{3}. 
 
     For  long term storage of solid $\CON{H}$ pellets,  passive  systems 
using permanent or superconducting magnets are probably the  most 
promising. If  some forces (due for example to the acceleration 
of  a rocket) are acting on the pellet,  the magnetic  levitation 
system may be aided by an electrostatic field or a laser beam  to 
balance them.

\begin{figure}
\begin{center}
\resizebox{8cm}{!}{ \includegraphics{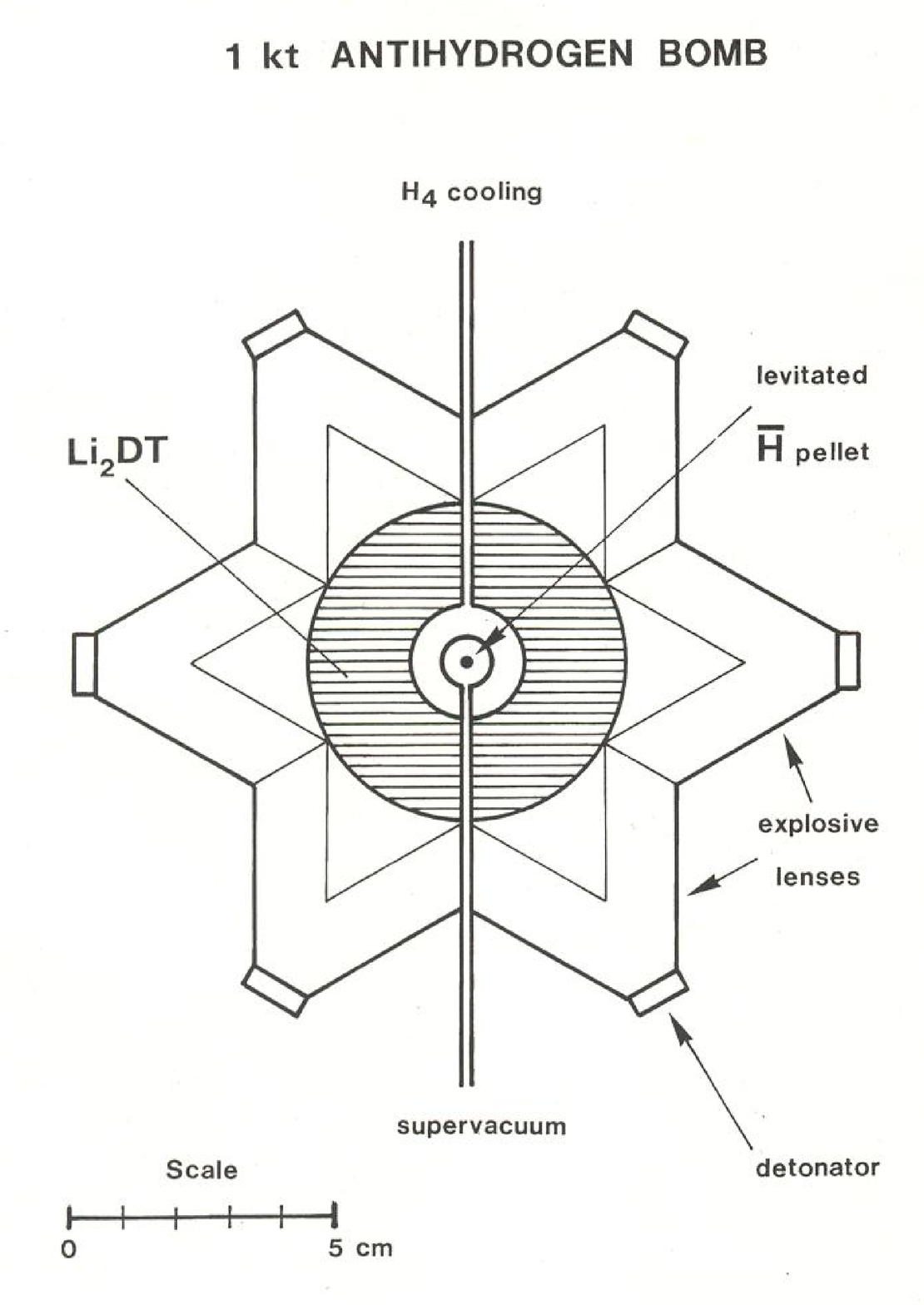}}
%..............................................
\end{center}
\caption[Antimatter bomb]{A possible design for a 1~kt  
antimatter bomb.   One microgram of 
antihydrogen  in a microcryostat is  levitated 
at  the  center  of  a  100~g  $Li_2DT$  sphere.  
Implosion  of the $Li_2DT$ by means  of  chemical 
explosives  brings the thermonuclear fuel into 
contact  with the  antihydrogen.   The  energy 
release  by  annihilation  is fast  enough  to 
trigger  an outgoing thermonuclear  detonation 
wave which burns the $Li_2DT$.   Depending on the 
amount   of   compression  by   the   chemical 
explosives,  the  device  operates as a  1~kt 
neutron bomb (ERW --- Enhanced Radiation Warhead) 
or a 1~kt blast bomb (RRR -- Reduced Residual radioactivity).  
In either case,  the antimatter bomb will have 
very   reduced   radioactive   fallout    and 
electromagnetic pulse effects.} 
\end{figure}

\section{Discussion and conclusions}
%----------------------------------- 

The  physics of matter-antimatter interaction has  been  reviewed 
and the main conclusions can be summarized as follows: 

\begin{itemize}
 
     \item   Plasma heating by the charged particles produced  in $\CON{p}$ 
annihilation  with  protons  or nuclei is  a  rather  inefficient 
process.    However,  if  the  fuel  to  be  heated  is  slightly 
compressed  ($\kappa=30$),  the  energy from the low velocity  particles 
(protons,  recoiling  nuclei) can be contained to give an  energy 
deposition of about 15 to 35~MeV per annihilation in $DT$ or  $Li_2DT$. 
 
     \item   Annihilation  in  a  hot  matter-antimatter  plasma  is 
relatively  slow:  about 5~ns for a $H-\CON{H}$ plasma.   However,  if  a 
small  amount  of  antimatter is brought in  contact  with  dense 
matter,  annihilation  in the boundary layer is  quite  fast.   A 
pellet  of  $\CON{H}$ disappears in about 0.5 ns or respectively  0.2~ns 
when it comes in contact with a collapsing shell of compressed $DT$ 
or  $Li_2DT$.   This  very  short  duration  energy  release  makes 
antimatter a good candidate as an energy source for pumping X-ray 
or gamma-ray lasers \cite{4}. 
 
     \item   The  number  of  $\CON{p}$ annihilations required  to  start  a 
thermonuclear  burn  wave in either $DT$ or  $Li_2DT$ is  found  to  be 
about  $10^{21}/k^2$.  Thus for  $\kappa=30$ (about the maximum compression 
factor that can be achieved with chemical explosives), $N = 10^{18}$. 

\end{itemize}
 
     The   technologies  for  producing  $\CON{p}$'s  with  high   energy 
accelerator  systems,  and the means for manipulating and storing 
sizable amounts of $\CON{H}$ have been examined.   With reference to  the 
conceptual  design of a 1~kt antihydrogen-triggered thermonuclear 
bomb  shown  in Fig.~5,  and with the objective of  designing  an 
antimatter  factory capable of producing the $10^{13}$ $\CON{p}$/s needed  for 
manufacturing  one such $\CON{H}$-bomb per day,  the main results can  be 
concluded as follows: 

\begin{itemize}
 
     \item   Under  ideal  conditions such  as  highly  efficient  $\CON{p}$ 
collection  and  very small $\CON{p}$ losses throughout  the  plant,  the 
production of $10^{13}$ $\CON{p}$/s requires either a fixed target system with 
a 1~mA, 120~GeV proton accelerator, or a (16+16)~GeV colliding beam 
$\CON{p}$ source with a proton supply current of 0.6~mA.   Assuming an AC 
power  to beam power efficiency of 25\%,  the accelerator's 
electric power requirement is 500~MW in the fixed target and about 
50~MW in the collider system. The energy needed to produce one 
$\CON{p}$  is thus of the order of $10^{-6}$~J, so that the production of  $10^{16}$  $\CON{p}$'s   for   each 
antimatter  triggered ICF pellet would require 
an energy investment of at least $10^4$~MJ.  
It will therefore be very difficult to achieve 
energy break-even in power generating  reactors 
using annihilation techniques.  Therefore,  civilian  
applications of antimatter for power production are very unlikely. 
 
     \item   If  the expected enhancement of $\CON{p}$ production in  high-energy
heavy-ion collisions is demonstrated by experiment, a high 
luminosity  heavy-ion collider would probably be the best  source of 
$\CON{p}$'s.  Compared with a fixed target $\CON{p}$ source, a collider can be 
designed to optimize conditions for $\CON{p}$ collection and cooling. 
 
     \item   The most difficult problem is the cooling of  the  very 
hot  $\CON{p}$'s  produced in high energy collisions to temperatures  low 
enough  so  that they can be permanently stored  in  (relatively) 
simple systems,  or combined with positrons to form antihydrogen.  
The  state-of-the-art is short by 2 to 4 orders of  magnitudes  from 
being able to cool $10^{13}$ $\CON{p}$/s.   However,  possible improvements in 
stochastic and electron cooling will probably bridge the gap. 
 
     \item   As  intense beams of cold low-energy  positrons  become 
available, $\CON{H}$ formation becomes easier. Small pellets of solid $\CON{H}$ 
can be levitated in a vacuum by a variety of  magnetic,  electric 
and  laser techniques,  and stored for very long periods if  the 
vacuum is better than $10^{-15}$~Torr. 
 
     \item   If  $\CON{p}$  cavitation  in  superfluid  helium is found 
experimentally,  formation of $\CON{H}$ would not be necessary for long 
term bulk storage of antimatter. 
 
     \item  The  electromagnetic levitation of a $10^{-6}$~g  $\CON{H}$  pellet 
within  a 1~mm diameter microcryostat at the center of a  large 
 $Li_2DT$  sphere  such as in Fig.~5 is a tremendous challenge  for 
materials microtechnology.  However,  if 
metastable  states  of  $\CON{p}$'s  in  $Li-$, $Be-$ or 
possibly $C-DT$ compounds are  discovered,  much 
simpler designs could be considered. 

\end{itemize}
 
     Before concluding, we note that a plant of the size required to produce  the antimatter  needed for one thermonuclear  bomb 
trigger a day ($10^{-6}$g of $\CON{H}$ or $10^{18}$ $\CON{H}$ atoms per 
day)  could consist of several 10's of accelerators 
and storage rings,  and could require 
as many as several large nuclear power  plants 
to  supply the electricity.   However,  
considering  the advances in  technology 
since 1945,  the relative complexity and cost of such a $\CON{p}$ factory 
are  not  out  of  proportion  with  those  of  a  large  uranium 
enrichment plant.   Indeed, a study  by 
the  RAND Corporation gives a cost estimate of 
\$500 to 1000~million  for  a  prototype  factory 
providing 10 to 100~micrograms, and \$5 to 15~billion 
for  a full production factory with an  output 
of about 10~mg per year \cite{4}.  

     From   the   point   of  view   of   
non-proliferation  of nuclear  weapons,  the  fact 
that     antimatter-triggered    thermonuclear 
weapons    will   have    extremely    reduced 
radioactive  fallout,  even for ground bursts, 
is  an important  consideration.   Since  such 
explosives  may  be  advocated  for  "peaceful 
nuclear   explosions,"    the   current   
non-proliferation regime is being threatened by the 
growing  spread  of  high  energy  accelerator 
technologies \cite{41}.  Moreover, from a strategic 
point   of  view,   the  possible  advent   of 
extremely   compact   and  essentially   clean  
nuclear  weapons  would  further  diffuse  the 
distinction between low-yield nuclear  weapons 
and conventional explosives.  Finally,   in  the  event  of  a 
comprehensive  test ban treaty,  antimatter would provide a means 
for   triggering   laboratory  and  small   scale   thermonuclear 
explosions  in  a yield range which cannot easily be  covered  by 
underground explosions or classical ICF systems \cite{41}.

\end{document}